\documentclass[nofootinbib,prd]{revtex4}
\pdfoutput=1

\begin{document}

\title{Quantization of spherically symmetric loop quantum gravity coupled to a scalar field  and a clock: the asymptotic limit}

\author{Rodolfo Gambini$^1$,  Jorge Pullin\footnote{Corresponding author. Email: pullin@lsu.edu}$^2$}
\affiliation{1. Instituto de F\'{\i}sica, Facultad de Ciencias, Igu\'a 4225, esq. Mataojo,
11400 Montevideo, Uruguay. \\
2. Department of Physics and Astronomy, Louisiana State University,
Baton Rouge, LA 70803-4001, USA.}

\begin{abstract}
We continue our work on the study of spherically symmetric loop quantum gravity coupled to two spherically symmetric scalar fields, one which acts as a clock. As a consequence of the presence of the latter, we can define a true Hamiltonian for the theory. The spherically symmetric context allows to carry out precise detailed calculations. Here we study the theory for regions of large values of the radial coordinate. This allows us to define in detail the vacuum of the theory and study its quantum states, yielding a quantum field theory on a quantum space time that makes contact with the usual treatment on classical space times.
\end{abstract}
\maketitle

\section{Introduction}
Spherically symmetric loop quantum gravity is a good symmetry reduced laboratory in which one can study black holes, singularity elimination by the quantum theory, and other issues, and has been developed over a decade by now \cite{usreview}. The introduction of matter has, however, proved problematic. In the vacuum theory one uses a redefinition of the constraints that allows to turn them into a Lie algebra and complete the Dirac quantization, which at present is not known to exist in the case coupled to matter at the quantum level. Being able to include massless scalar fields is a potentially attractive setting, as it is known to have a rich dynamics that  includes black hole formation and also the critical phenomena discovered by Choptuik \cite{choptuik}. 

Here we would like to expand on our previous paper \cite{previous} that considered a spherically symmetric massless scalar field coupled to spherically symmetric gravity in the presence of a clock given by a second scalar field. The latter gives rise to a true Hamiltonian so one is quantizing a gauge fixed theory and does not have to worry about constraints. This avenue of  using  matter clocks in quantum gravity has been considered by other authors as well (see \cite{gravityquantized} for references). In our approach we  exploit the advantages of the simplifications due to spherically symmetric gravity to make progress in defining the relevant quantum operators in a precise way. In this paper we will consider using approximations to carry out concrete calculations of the space of states of the coupled theory. We will consider the theory for large values of the radial coordinate and expand it in powers of Newton's constant. Since we are in spherical symmetry, that would mean far away from a black hole or star. This makes calculations considerably more tractable. We will be able to show that in the asymptotic region the theory admits a large family of quantum vacua for quantum matter fields coupled to quantum gravity, as it is expected from the well known results of quantum field theory on classical curved space time.

The organization of this article is as follows: in section 2 we review the classical theory. In section 3 we consider the eigenvalue problem of the  gravitational Hamiltonian and we study the spectrum in the vacuum case.
In section 4 we introduce a perturbative treatment of matter on a quantum gravitational space time and analyze the allowed fundamental states of the coupled system in the asymptotic region. We end with conclusions.

\section{Classical theory: spherical gravity with a scalar field and a  clock}

Following the usual treatment \cite{usreview}, in spherically symmetric loop quantum gravity one has as canonical variables the triads in the radial and tangential directions $E^x, E^\varphi$ and their canonical momenta $K_x,K_\varphi$. The variables for the scalar field are $\phi$ and $P_\phi$, and for the second scalar field we use as a clock $\psi$ and $P_\psi$ (we take $\hbar=c=1$). 

The total constraint for the theory is,
\begin{eqnarray}
    H_T&=&\frac{1}{G}\int dx \left[ N^x\left(\left(E^x\right)'K_x-E^\varphi\left(K_\varphi\right)'-8\pi G P_\phi \phi'-8\pi G P_\psi \psi'\right)\right.\nonumber\cr
    &&+N
    \left(
    -\frac{E^\varphi}{2\sqrt{E^x}}-2\sqrt{E^x}K_\varphi K_x-\frac{K_\varphi^2 E^\varphi}{2\sqrt{E^x}}
    +\frac{\left(\left(E^x\right)'\right)^2}{8\sqrt{E^x}E^\varphi}
    -\frac{\sqrt{E^x}\left(E^x\right)'\left(E^\varphi\right)'}{2\left(E^\varphi\right)^2}\right.\nonumber\cr
    &&\left.\left.+\frac{\sqrt{E^x}\left(E^x\right)''}{2 E^\varphi} +\frac{2\pi G P_\phi^2}{\sqrt{E^x}E^\varphi}
    +\frac{2\pi G\sqrt{E^x}{E^x}\left(\phi'\right)^2}{E^\varphi}
    +\frac{2\pi G P_\psi^2}{\sqrt{E^x}E^\varphi}
    +\frac{2\pi G\sqrt{E^x}{E^x}\left(\psi'\right)^2}{E^\varphi}
    \right)\right].
\end{eqnarray}
As discussed in our previous paper, we redefine the shift,
\begin{equation}
    N^x=N^r+\frac{2 N \sqrt{E^x}K_\varphi}{\left(E^x\right)'},
\end{equation}
demand that $E^x$ is a time independent function, which fixes $N^x=0$ and $K_x$.

The gravitational part can be written  in terms of the  derivative of $C$, given by,
\begin{equation}
    C=-\sqrt{E^x}\left(1+K_\varphi^2 -\frac{\left(\left(E^x\right)'\right)^2}{4\left(E^\varphi\right)^2}\right) +2GM,
\end{equation}
 where the last term arises from the boundary condition \cite{kuchar,rastgoo}.
 The vanishing of the Hamiltonian constraint implies that $C$ (and its derivative) vanish in vacuum. Since we will be working in the asymptotic region, our treatment would be valid for black hole space-times and other spherically symmetric space-times.

The Hamiltonian constraint for the system  written in terms of the re-scaled lapse $N_{\rm new}$ can be written as,
\begin{eqnarray}
H_{\rm total} &=&
\frac{2 N_{\rm new} \pi}{\sqrt{E^x}\left(E^\varphi\right)^2}
\left[\left(E^x\right)'\left(
\left(E^x\right)^2\left(
 \left(\phi'\right)^2+
 \left(\psi'\right)^2
\right)
+P_\phi^2+P_\psi^2
\right)
-8E^x E^\varphi K_\varphi 
\left(\phi'P_\phi+\psi' P_\psi\right)
\right]+\frac{N_{\rm new}}{G} C'.\label{17}
\end{eqnarray}

As in the previous paper, we now fix the gauge. We do so by choosing the clock $\phi=t/l_0^2$ with $l_0$ a constant with units  of length; recall that we work in units where $\hbar=c=1$, and impose that the gauge fixing is preserved upon evolution. $l_0$ is a parameter that determines the size of the spatial region 
for which the clock chosen determines simultaneity and $t$ is the asymptotic proper time. 
In addition to that we will later choose $E^x$ as a given function of $x$. This will determines the lapse as , 
\begin{equation}
    N_{\rm new}=\frac{\sqrt{E^x} \left(E^\varphi\right)^2}{4 \pi P_\phi \left(E^x\right)' l_0^2}.\label{18}
\end{equation}

The Hamiltonian constraint and the gauge fixing of the clock field become now second class equations. They
allow  to impose strongly the constraint in the sense of Dirac and to obtain an equation for $P_\phi$ and $\phi$.  The Dirac Brackets for $\psi,P_\psi,K_\varphi,E^\varphi$ coincide with the Poisson Brackets.

We can write the evolution equations for the remaining variables,  $K_\varphi, E^\varphi, \psi$ and $P_\psi$ by considering their Poisson brackets with the true Hamiltonian,
\begin{equation}
    H_{\rm true}=M+\int_{-\infty}^\infty dx
     \frac{ \sqrt{
    -C' \sqrt{E^x}\left(E^\varphi\right)^2
    -2\pi G
    \left(
    \left(E^x\right)'\left(E^x\right)^2\left(\psi'\right)^2
    -8E^x K_\varphi E^\varphi P_\psi \psi'+\left(E^x\right)'P_\psi^2
    \right)}}
    {l_0^2 \sqrt{2\pi G \left(E^x\right)'}}.
\end{equation}

The Hamiltonian constraint of the complete theory ensures that the quantity under the square root is always positive in the classical theory, since in the asymptotic region considered it is a positive quantity times $P_\phi^2$. At a quantum level the presence of the square root requires an  extension of the operator, so it is well defined on the complete phase space including the portions excluded classically. This is achieved by taking the absolute value of the argument of the square root \cite{GieselThiemann}. 

\section{Perturbative quantum treatment: the vacuum case}

In preparation to treat the quantum theory, it is convenient to absorb the part that diverges in the continuum limit of the portion of the Hamiltonian of the scalar field. In full four dimensions that would be a term akin to a cosmological constant. In spherical symmetry, it is tantamount to a deficit angle (see for instance \cite{rastgoo}), we extract the divergent part of the matter field in the continuum limit, $\rho_{\rm vac}$ and we add it to the gravitational part through the introduction of the deficit angle $\Lambda$,
\begin{equation}
    H_{\rm true}=M+\int_{-\infty}^\infty dx
     \frac{ \sqrt{
    -C' \sqrt{E^x}\left(E^\varphi\right)^2
    -2\pi G
    \left(
    \left(E^x\right)'\left(E^x\right)^2\left(\psi'\right)^2
    -8E^x K_\varphi E^\varphi P_\psi \psi'+\left(E^x\right)'P_\psi^2
    -\rho_{\rm vac} \left(E^\varphi\right)^2 \sqrt{E^x}
    \right)}}
    {l_0^2 \sqrt{2\pi G \left(E^x\right)'}}.
\end{equation}

One sees that the gravitational and matter part contributions to the quantity in the square root are opposite signs, we will see that plays an important role. We subtract the $\rho_{\rm vac}$ term to ensure that the matter term is much smaller than the gravitational part and we keep the true Hamiltonian unchanged by introducing a deficit angle in the gravitational part of the contribution. When quantizing we need to ensure the self adjointess of the Hamiltonian, which implies that the expression under the square root should be positive. This may be accomplished by considering its absolute value. This does not modify the classical solutions if one takes into account the Hamiltonian constraint. 

Let us consider the dominant term in the vacuum limit $G=0$. As usual in spherically symmetric loop quantum gravity one takes a kinematical basis of quantum eigenstates of the operators $\hat{E}^x$ and $\hat{E}^\varphi$ obtained by the direct product of a one dimensional loop representation along a graph in the radial direction times a Bohr compactification in the transverse direction. That is,
\begin{eqnarray}
\hat{E}^x_j \vert k_1,\ldots, k_N,\mu_1,\ldots, \mu_n\rangle&=&k_j \ell_{\rm Planck}^2\vert k_1,\ldots, k_n,\mu_1,\ldots, \mu_n\rangle,\\
\hat{E}^\varphi_j \vert k_1,\ldots, k_n,\mu_1,\ldots,\mu_n\rangle&=&\mu_j \ell_{\rm Planck}\vert k_1,\ldots, k_N,\mu_1,\ldots, \mu_n\rangle.
\end{eqnarray}

We consider a proper basis of states for the gravitational part of the Hamiltonian constraint $C_j$,
\begin{equation}
    \hat{C}_j=-j \Delta \left(1-2 \Lambda +\frac{\sin\left(\rho \hat{K}_{\varphi,j} \right)^2}{\rho^2} -\frac{\left(\hat{E}^x_{j+1}-\hat{E}^x_j\right)^2}{4 \Delta^2 \left(\hat{E}^\varphi_j\right)^2}\right) + 2GM,
\end{equation}
where $\rho$ is the polymerization parameter of the Bohr compactification, $\Lambda=2 \pi G \rho_{\rm vac}$. The operator $\hat{E}^x_j$ commutes with $\hat{H}_{\rm true}$ and therefore is a constant of the motion that in the spin network representation has eigenvalues $k_j\ell_{\rm Planck}^2$ with $k_j$ integers. In order to simplify things we choose an equally-spaced lattice with $E^x(x)=x^2$,  $x_j
=j \Delta+x_0 $, with $\Delta=n \ell_{\rm Planck}$ the lattice spacing and $k_j={x_j}^2/\ell_{\rm Planck}^2$, $n$ an integer and $x_0\gg r_S=2GM$ so we are in the asymptotic region. To study the quantity under the square root in the vacuum case, We consider the eigenvalue equation, 
\begin{equation}
    \hat{C}_j \psi_{l_j}(\mu_j)=
    l_j \psi_{l_j}(\mu_j),
\end{equation}
with $j=1,\ldots,N$. Because $\hat{C}_j$ acts only at $j$, we introduce the states $\psi_{l_j}(\mu_j)$. One can construct states of the kinematical basis we considered before by taking their direct product.

We need the inverse of $\left(\hat{E}^\varphi\right)^2$, and using $E^\varphi\vert \mu\rangle=\ell_{\rm Planck}\mu\vert \mu\rangle$ (we are dropping the index $j$ in $\hat{E}^\varphi$ and $\mu$ to simplify the notation) and the Thiemann trick, we have that,
\begin{equation}
\frac{(2x)^2}{4 \ell_{\rm Planck}^2}\left(\hat{E}^\varphi\right)^{-2} \vert\mu\rangle= \frac{729 x^2}{4096 \ell_{\rm Planck}^2 \rho^6} \left(\left(\mu+\rho\right)^{2/3}-\left(\mu-\rho\right)^{2/3}\right)^6\vert \mu\rangle, 
\end{equation}
which in the limit of large $\mu$'s, that is $\rho/ \mu \ll 1$ gives for the eigenvalue $\left(\hat{E}^\varphi\right)^{-1}=1/\left(\ell_{\rm Planck} \mu\right)$.

The asymptotic approximation we are using also  allows us to approximate the eigenvalue equation as a differential equation that for small $\rho$ and sufficiently large values of $\mu$ (as one has in the asymptotic region, as we shall see), takes the form, recalling the gauge choice $E^x=x^2$,
\begin{equation}
\frac{d^2}{d\mu^2} \psi_l(\mu)-l_x \psi_l(\mu)+\frac{x^2}{\ell_{\rm Planck}^2\mu^2} \psi_l(\mu)=0,
\end{equation}
with,
\begin{equation}
    l_x=\frac{l}{x}-\frac{r_S}{x} +1 -2 \Lambda.
\end{equation}
We need to require  $l_x>0$ in order to have normalizable solutions, and therefore $l>2 \Lambda x -x+r_S$. 

We now introduce a redefinition $\psi_l(\mu)=\sqrt{\mu}\phi_l(\mu)$, which allows to rewrite the eigenvalue equation as,
\begin{equation}
    \mu^2 \frac{d^2}{d\mu^2}\phi_l(\mu) +\mu \frac{d}{d\mu} \phi_l(\mu) -\phi_l(\mu)\left( \frac{1}{4} +\mu^2 l_x  - \frac{x^2}{\ell_{\rm Planck}^2}\right)=0.
\end{equation}

The redefinition of the variable helps us deal with the fact that the original eigenvalue equation is ill defined for $\mu=0$. 
We are interested in studying the field in the asymptotic region $x\gg r_S$ where we can neglect the $\phi/4$ term with regards to the $x^2/\ell_{\rm Planck}^2$ one.
Introducing $\alpha=\mu \sqrt{(l_x)}\ell_{\rm Planck}/x$, we can rewrite,
\begin{equation}
    \frac{\alpha^2 \ell_{\rm Planck}^2}{x^2}\frac{d^2}{d\alpha^2}\phi(\alpha)
    +\frac{\alpha\ell_{\rm Planck}^2}{x^2} \frac{d}{d\alpha}\phi(\alpha)+\left(1-\alpha^2\right)\phi(\alpha)=0,\label{15}
\end{equation}
which has as solutions,
\begin{equation}
    \phi(\alpha)=C_1 {\rm I}\left(\frac{i x}{\ell_{\rm Planck}}, \frac{\alpha x}{\ell_{\rm Planck}}\right) +C_2  {\rm K}\left(\frac{{ix}}{ \ell_{\rm Planck}}, \frac{\alpha x }{\ell_{\rm Planck}}\right), 
\end{equation}
with $I,K$ are modified Bessel functions of the first and second kind,
which, given that $\phi =c \psi /\alpha^{1/4}$, allows us to write,
\begin{equation}
    \psi(\alpha)=A\, {\alpha}^{1/4}{\rm I}
    \left(\frac{ix}{\ell_{\rm Planck}}, \frac{\alpha x }{\ell_{\rm Planck}}\right)
    +B\, {\alpha}^{1/4}{\rm K}
    \left(\frac{ix}{\ell_{\rm Planck}}, \frac{\alpha x }{\ell_{\rm Planck}}\right).
\end{equation}

These solutions, with appropriate choices of the constants are approximated for large $x\gg r_S$ by Dirac deltas, they are proportional to $\delta(\alpha-1)$ and $\delta(\alpha+1)$. For simplicity, for now on, we will use the approximate expressions. This can be easily seen from equation (\ref{15}) since the first two terms are very small in the asymptotic region for normalizable solutions. As is usual in quantum mechanics we need to eliminate solutions that diverge at infinity to determine the eigenstates.
The solutions for $\alpha$ positive and negative are independent of each other, so we will label the elements of the space of solutions of $\hat{C}$ with $\pm$ indicating the corresponding sign of $\alpha$.

Recalling that $\alpha$ is a function of $\mu$, the eigenfunctions of $\hat{C}$ can be rewritten as,
\begin{eqnarray}
    \langle \mu\vert l, +\rangle&=&
    \frac{\sqrt{2}}{2\sqrt{\ell_{\rm Planck}}l_x^{3/4}}
    {\delta\left(\mu-\frac{x}{\ell_{\rm Planck}\sqrt{l_x}}\right)}{},\\    \langle \mu\vert l, -\rangle&=&
    \frac{\sqrt{2}}{2\sqrt{\ell_{\rm Planck}}l_x^{3/4}}
    {\delta\left(\mu+\frac{x}{\ell_{\rm Planck} \sqrt{l_x}}\right)},
\end{eqnarray}
and this constitutes a basis of states with a continuous spectrum.

And  in the asymptotic region  we have that,
$
\langle l,+\vert l',+\rangle=\delta(l-l')$ and similarly for the $\vert l,-\rangle$ and $\langle l,+\vert l,-\rangle=0$, which completes the diagonalization of the operator associated to the Hamiltonian constraint
$\hat{C}$. As a consequence, these relations can be summarized as $\langle l,\epsilon\vert l,\epsilon'\rangle=\delta_{\epsilon,\epsilon'}\delta(l-l')\rangle$ with the first a Kronecker delta. One can also show that the basis $\vert l,\epsilon \rangle$ is complete, that is,
\begin{equation}\int dl \sum_\epsilon \vert l, \epsilon\rangle \langle l,\epsilon\vert =1.
\end{equation}

We can now determine the form of the eigenvalue equation of the gravitational Hamiltonian, that as we saw is given by,
\begin{equation}
H_{\rm grav} =\int_{-\infty}^\infty
\frac{\vert E^\varphi(x)\vert \sqrt{\vert C'(x)}\vert \sqrt{\vert E^x(x)\vert}}{ \sqrt{2\pi G (E^x(x))'} l_0^2}.
\end{equation}
Introducing $F=\left(\sqrt{2 \pi} \ell_{\rm Planck} l_0^2\right)^{-1}$ and recalling that the lattice spacing is $\Delta$, we have that,
\begin{equation}
\hat{H}_{\rm grav} =\sum_i \frac{F \Delta}{2\sqrt{2}}
 \left( \vert \hat{E}^\varphi_i\vert 
 \sqrt{ \frac{\vert\hat{C}_{i+1}-\hat{C}_i\vert}{\Delta}}
+\sqrt{\frac{\vert\hat{C}_{i+1}-\hat{C}_i\vert}{\Delta}}\vert \hat{E}^\varphi_i\vert\right).
\end{equation}

We wish to study the eigenvalue equation for this operator in the asymptotic region $x\gg r_S$, that is,

\begin{equation}\label{23}
\hat{H}_{\rm grav}
\Psi\left(l_1,\ldots, l_N\right) = E
\Psi\left(l_1,\ldots, l_N\right),
\end{equation}

Its explicit form is 
\begin{eqnarray}
    \hat{H}_{\rm grav}
\Psi&=&
\sum_i \int dl' \int d\mu\left[
\frac{F\Delta}{2}
\left\vert\left(\psi^i\right)^*_{l,\epsilon}\left(\mu\right) \ell_{\rm Planck}\right\vert \mu\left\vert \psi^i_{l',\epsilon}\left(\mu\right)\right\vert 
\left(\sqrt{\frac{1}{2\Delta}\left\vert l_{i+1}-l'_i\right\vert}\right)\Psi\left(l'_i,l_{i+1}\right)
\right.\nonumber\\
&&\left. +
\frac{F\Delta}{2}
\left(\sqrt{\frac{1}{2\Delta}\left\vert l_{i+1}-l_i\right\vert}\right)
\left(\psi^i\right)^*_{l,\epsilon}\left(\mu\right) \ell_{\rm Planck}\vert \mu\vert \psi^i_{l',\epsilon}\left(\mu\right) 
\Psi\left(l'_i,l_{i+1}\right)\right],
\end{eqnarray}
and we recall that the operator $\hat{E}^\varphi$ is diagonal in the $\mu$ basis and 
$
\psi^i_{l',\epsilon}\left(\mu\right)=
\langle \mu\vert l'_i,\epsilon\rangle
$ and 
$
\left(\psi^i\right)_{l',\epsilon}^*\left(\mu\right)=
\langle l'_i,\epsilon\vert \mu\rangle
$ and the index $i$ refers to which $\mu$ it is acting on.

Taking into account that
\begin{equation}
\langle l,+ \vert \left(\vert\mu\vert \right)\vert l',+\rangle =\frac{x}{\ell_{\rm Planck}\sqrt{l_x}} \delta(l'-l),
\end{equation}
which can be rewritten as
\begin{equation}
    \hat{H}_{\rm grav}
\Psi\left( l_1,\ldots,   l_N,M\right)
=\sum_j
\frac{F \sqrt{\Delta} \sqrt{2}}{\sqrt{l_{x_j}}}
x_j \sqrt{\vert  l_{j+1}-l_j\vert}\Psi\left(l_1,\ldots,l_j,l_{j+1},\ldots,l_N,M\right).
\end{equation}

Equation (\ref{23}) can be immediately solved. Its solution for the fundamental state takes the form,
\begin{equation}
    \Psi(l_1,\ldots,l_N,M)=\delta(l_2-l_1)\delta(l_3-l_2)\cdots \delta(l_N-l_{N-1})\delta(M-M_0),
\end{equation}
for a space-time of a black hole of mass $M_0$.

Which can be rewritten as,
\begin{equation}
    \Psi(l_1,\ldots,l_N,M)=\delta(l_2-l_1)\delta(l_3-l_1)\cdots \delta(l_N-l_1)\delta(M-M_0),
\end{equation}
which ends up being more convenient later.

This solution is an element of a continuous spectrum and is therefore not normalizable, as one encounters in quantum mechanics for the free particle. If one wishes to compute probabilities one needs to construct wave packets that are normalizable, and in exact analogy with the free particle take the form,
\begin{equation}
\psi(p)=\left(\frac{\sigma}{\pi}\right)^{1/4}\exp\left(-\frac{p^2\sigma}{2}\right),
\end{equation}
where $\sigma$ is the inverse of the dispersion of the Gaussian and we take it to be a large positive quantity so its square approximates the Dirac delta. Therefore the wavefunction for the states with lower energy is,
\begin{equation}
    \Psi_{\rm grav}\left(l_1,\ldots,l_{N},M\right)=\left(\frac{\sigma}{\pi}\right)^{N/4}
    \exp\left(-
    \sum_k \frac{\left(l_k-l_1\right)^2\sigma}{2}
    \right)\delta(M-M_0).
\end{equation}

\section{Perturbative quantum treatment: inclusion of matter}

We will now expand the Hamiltonian $H_{\rm true}$

\begin{equation}
    H_{\rm true}=\frac{E^\varphi}{4\sqrt{\pi G} l_0^2}
    \sqrt{
    -8\pi G \left(\phi'\right)^2x^4 \left(E^\varphi\right)^{-2}
    +32\pi G \phi' K_\varphi P_\phi x\left(E^\varphi\right)^{-1}
    -8\pi G P_\phi^2\left(E^\varphi\right)^{-2}
    +4\pi G \rho_{rm vac} -2 C'
    },
\end{equation}
in powers of $\sqrt{G}/l_0$, which yields, keeping terms of order one,
\begin{eqnarray}
    H_{\rm true}&=&H_{\rm grav}+H_{\rm matt}=
\frac{    \vert E^\varphi\vert \sqrt{-2 C'}}
{4 \sqrt{\pi G}\,l_0^2}\nonumber\\
&&
-\frac{
\vert E^\varphi\vert 
\sqrt{\pi G}\sqrt{-2 C'}
\left(
2 \left(\phi'\right)^2 x^4-8 E^\varphi \phi' K_\varphi P_\phi x+\left(E^\varphi\right)^2 \rho_{\rm vac}+2 P_\phi^2
\right)}
{
4 \left(E^\varphi\right)^2 \left(E^x\right)' (-C') l_0^2
}.\label{31}
\end{eqnarray}

The dominant order term only depends on the gravitational variables and is the one we studied in the previous section. The first order correction in powers of $\sqrt{G}$ has matter and gravitational variables. The latter will be substituted by their expectation values as discussed in the previous section. We are therefore working in the approximation of quantum field theory in a quantum space time. 

Let us compute the relevant expectation values. For that, we consider the state we  discussed at the end of the last section
\begin{equation}
    \Psi_{\rm grav}\left(l_1,\ldots,l_{N},M\right)=\left(\frac{\sigma}{\pi}\right)^{N/4}
    \exp\left(-
    \sum_k \frac{l_k^2\sigma}{2}
    \right),
\end{equation}
where we have chosen $l_1=0$, which corresponds to an interior boundary condition since we are working in a finite domain. For large $\sigma$'s, this approximates the fundamental state given by the product of the Dirac deltas.

To compute the expectation values of the variables at a point, we need the gravitational state written as a function of $\mu_j$,
\begin{equation}
    \Psi_{\rm grav}\left(\mu_j\right)
    =
    \sqrt{2 \ell_{\rm Planck}}
    \left(\frac{\sigma}{\pi}\right)^{1/4}
    \left(\frac{ x_j^2}{\ell_{\rm Planck}^2 \mu_j^2}
    \right)^{3/4}
    \exp\left(
    -\left[
2x_j\Lambda+\frac{x_j^3}{\ell_{\rm Planck}^2\mu_j^2}-{x_j+r_S}
    \right]^2 \frac{\sigma}{2}
    \right).
\end{equation}

Therefore the expectation value of  $\hat{E}^\varphi_k$ is given by,
\begin{equation}
\langle \Psi\vert \left(\vert \hat{E}^\varphi_k\vert\right)\vert \Psi\rangle= 
\left\langle \Psi\left\vert 
\mu_k\ell_{\rm Planck}
\right\vert \mu_k\right\rangle\left\langle \mu_k
\right\vert\left. \Psi\right\rangle
=\frac{x_k}{\sqrt{1-2\Lambda -\frac{r_S}{x_k}}}.
\end{equation}

Its dispersion is given by 
\begin{equation}
\langle \Psi\vert \left(\hat{E}_k^\varphi\right)^2\vert \Psi\rangle-
\langle \Psi\vert
\hat{E}_k^\varphi\vert \Psi\rangle^2
=
    {\frac {1}{16\sigma}\sqrt {-{\frac { \left( 128\,\sigma\, \left( 
\Lambda-1/2 \right) ^{2}{x_{{k}}}^{2}+128\,{\it r_S}\,\sigma\, \left( 
\Lambda-1/2 \right) x_{{k}}+32\,{{\it r_S}}^{2}\sigma-9 \right) {x_{{k}
}}^{3}}{ \left(  \left( 2\,\Lambda-1 \right) x_{{k}}+{\it r_S} \right) 
^{5}}}}}.
\end{equation}

In the asymptotic region we are considering $\sin(\rho \hat{K}_\varphi)/\rho$ can be approximated  $\hat{K}_\varphi$. Its expectation value vanishes. One computes it by recalling that $\hat{K}_\varphi$ is given by the derivative of the state with respect to $\mu$. Its integral between $\mu=\pm \infty$ therefore vanishes automatically. One can also compute the dispersion by taking the second derivative and it grows with $\sigma$. The uncertainty in $\hat{E}^\varphi$ diminishes with $\sigma$, as one would expect for variables that are canonically conjugate.
The expectation value of the product of $\hat{K}_\varphi$ and $\hat{E}^\varphi$ also vanishes for large $\sigma$'s. 

We now proceed to compute the expectation value of the gravitational and matter portions of the true Hamiltonian. In order for the expressions to make contact with the usual ones of quantum field theory on flat space-time, it is convenient to choose the width of the Gaussians in the following way,
\begin{equation}
\sigma_k=\frac{1}{\Delta^2}\left(\frac{l_0^a\Delta^{1-a}}{x_k}\right)^4.
\end{equation}
This choice of width of the Gaussians depends on a parameter $a$. The physical relevance of this parameter will become clear later. For simplicity from now on we will list expressions for the case $a=1$, and we will at the end discuss what changes for other values of $a$.

In loop quantum gravity one needs to polymerize the scalar field \cite{kaminski}. However, since we will be considering states where the scalar field is small, the polymerization can be ignored.

Substituting the expectation values of the gravitational variables in (\ref{31}) we get, for the true Hamiltonian density,
\begin{equation}
    \hat{H}_{{\rm true},j}=
    \frac{
    x_j^2\sqrt{2}
    }
    {4 \pi^{3/2} l_0^3\sqrt{G}}\sqrt{\frac{1}{1-2\Lambda}}
-\frac{\sqrt{2G}\pi^{3/2}}{l_0}\sqrt{1-2\Lambda}
\left[
\frac{\left(x_j \left(\hat{\phi}_{j+1}-\hat{\phi}_j\right)\right)^2}{2\Delta^2}
+\frac{\hat{P}_{\phi,j}^2}{2 \Delta^2 x_j^2}-\frac{\rho_{\rm vac}}{2-4\Lambda}
\right].
\end{equation}

And summing on $j$ we get the true Hamiltonian,
\begin{equation}
    \hat{H}_{\rm true}=
    \frac{\sqrt{2}}{12\sqrt{1-2\Lambda}\pi^{3/2}\sqrt{G}}-
    \Delta \sum_{j=0}^{j_N}
\frac{\sqrt{2G}\pi^{3/2}\sqrt{1-2\Lambda}}{l_0}
\left[
\frac{\left(x_j \left(\hat{\phi}_{j+1}-\hat{\phi}_j\right)\right)^2}{2\Delta^2}
+\frac{\hat{P}_{\phi,j}^2}{2 \Delta^2 x_j^2}-\frac{\rho_{\rm vac}}{2-4\Lambda}
\right],
     \label{38}
\end{equation}
and one sees that one has the difference of the gravitational and the matter portions. The first summation can be done by going to the continuum limit and computing the integral.

It is worth discussing briefly the extent of the asymptotic domain we have been considering. We choose $j_N=(l_0-x_0)/\Delta\sim l_0/\Delta$ assuming that the clock provides a good description of the complete asymptotic region considered. Notice that this also implies we are taking a large asymptotic region $[x_0,x_N]$ with $x_0\ll x_N$, and $x_0$ is the point closer to the black hole and $x_N$ the farthest.

The fields satisfy discrete canonical commutation relations $\left[\phi_k,P_{\phi,j}\right]=i \hbar \delta_{j,k}$. The middle term of (\ref{38}) is the Hamiltonian of a scalar field in a lattice of the radial direction of a flat space-time in spherical coordinates with a deficit angle.

We can write the equation of motion for the scalar field stemming from the above true Hamiltonian,
\begin{equation}
\ddot{\phi}_j
    x_j^{2}- \left( {\frac {\phi_{{j+1}
}-\phi_{{j}}}{\Delta}}-{\frac {\phi_{{j}}-\phi_{{j-1}}}{\Delta}}
 \right) x_j^{2}- \left( \,\phi_{{j}}-\,\phi_{{j-1}} \right) \left(2x
_{{j}}-\Delta\right)
=0,
\end{equation}
which can be approximately solved, being careful of not neglecting terms proportional to $\ell_{\rm Planck}$ times factors that could be large in the ultraviolet regime,
\begin{equation}
   \phi_{n,j} ={\frac { \left( a_{{n}}\exp\left({-i\omega_{{n}}t}\right)+{\it a^\dagger}_{{n}}
\exp\left({i\omega_{{n}}t}\right) \right) \sin \left( k_{{n}}{\overline x}_{{j}} \right) 
}{\sqrt {\pi\,k_{{n}}} {x}_{{j}}}},
\end{equation} 
where ${\overline x}_j=x_j-x_0$, $k_n=\frac{2 \pi n}{j_N\Delta}$ and $a_n$ and $a^\dagger_n$ will become annihilation and creation operators
and where $\omega$ is given by the typical dispersion relation in the lattice,
\begin{equation}
   \omega_n= \,{\frac {2 -2\cos \left( k_{{n}}\Delta \right) }{{\Delta}^{2}}}.
\end{equation}

We can now write the Hamiltonian density in terms of  creation and annihilation operators omitting the line over $x_j$,
\begin{eqnarray}
\hat{H}_{j}&=&\sum_{n=1}^N\left[
-1/2\,{\frac { \left( \sin \left( k_{{n}}x_{{j}} \right)  \right) ^{2}
\Delta\,{a_{{n}}}^{2}{\omega_{{n}}}^{2}}{\pi\,k_{{n}} \left( {{\rm e}^
{i\omega_{{n}}t}} \right) ^{2}}}
+{\frac { \left( \sin \left( k_{{n}}x_
{{j}} \right)  \right) ^{2}\Delta\,a_{{n}}{\omega_{{n}}}^{2}{\it a^\dagger}_{
{n}}}{\pi\,k_{{n}}}}
-1/2\,{\frac { \left( \sin \left( k_{{n}}x_{{j}}
 \right)  \right) ^{2}\Delta\,{{\it a^\dagger}_{{n}}}^{2}{\omega_{{n}}}^{2}
 \left( {{\rm e}^{i\omega_{{n}}t}} \right) ^{2}}{\pi\,k_{{n}}}}\right.
\nonumber\\
 &&
 +1/2\,{
\frac {{x_{{j}}}^{2} \left( \sin \left( k_{{n}}x_{{j+1}} \right) 
 \right) ^{2}{a_{{n}}}^{2}}{\Delta\,\pi\,k_{{n}}{x_{{j+1}}}^{2}
 \left( {{\rm e}^{i\omega_{{n}}t}} \right) ^{2}}}
 +{\frac {{x_{{j}}}^{2
} \left( \sin \left( k_{{n}}x_{{j+1}} \right)  \right) ^{2}a_{{n}}{
\it a^\dagger}_{{n}}}{\Delta\,\pi\,k_{{n}}{x_{{j+1}}}^{2}}}
-{\frac {x_{{j}}
\sin \left( k_{{n}}x_{{j+1}} \right) {a_{{n}}}^{2}\sin \left( k_{{n}}x
_{{j}} \right) }{\Delta\,\pi\,k_{{n}}x_{{j+1}} \left( {{\rm e}^{i
\omega_{{n}}t}} \right) ^{2}}}
\nonumber\\
&&
-2\,{\frac {x_{{j}}\sin \left( k_{{n}}x_
{{j+1}} \right) a_{{n}}\sin \left( k_{{n}}x_{{j}} \right) {\it a^\dagger}_{{n
}}}{\Delta\,\pi\,k_{{n}}x_{{j+1}}}}+1/2\,{\frac {{x_{{j}}}^{2} \left( 
\sin \left( k_{{n}}x_{{j+1}} \right)  \right) ^{2}{{\it a^\dagger}_{{n}}}^{2}
 \left( {{\rm e}^{i\omega_{{n}}t}} \right) ^{2}}{\Delta\,\pi\,k_{{n}}{
x_{{j+1}}}^{2}}}
\nonumber\\
&&
-{\frac {x_{{j}}\sin \left( k_{{n}}x_{{j+1}} \right) {
{\it a^\dagger}_{{n}}}^{2} \left( {{\rm e}^{i\omega_{{n}}t}} \right) ^{2}\sin
 \left( k_{{n}}x_{{j}} \right) }{\Delta\,\pi\,k_{{n}}x_{{j+1}}}}
 +1/2\,
{\frac { \left( \sin \left( k_{{n}}x_{{j}} \right)  \right) ^{2}{a_{{n
}}}^{2}}{\Delta\,\pi\,k_{{n}} \left( {{\rm e}^{i\omega_{{n}}t}}
 \right) ^{2}}}
 +{\frac { \left( \sin \left( k_{{n}}x_{{j}} \right) 
 \right) ^{2}a_{{n}}{\it a^\dagger}_{{n}}}{\Delta\,\pi\,k_{{n}}}}
 \nonumber\\
 &&\left.+1/2\,{
\frac { \left( \sin \left( k_{{n}}x_{{j}} \right)  \right) ^{2}{{\it 
a^\dagger}_{{n}}}^{2} \left( {{\rm e}^{i\omega_{{n}}t}} \right) ^{2}}{\Delta
\,\pi\,k_{{n}}}}\right].
\end{eqnarray}

The standard vacuum $\vert 0\rangle$ of the matter part is defined in the usual way, as the state on which the annihilation operators vanish. We can compute the expectation value of the Hamiltonian density in the vacuum, and we substituted $k_n$ in terms of $n$ using the form of $k_n$ given below (41),
\begin{eqnarray}
    \langle 0\vert \hat{H}_{j}\vert 0 \rangle
    &=&\sum_{n=1}^{j_N} n\frac{ \left(\left(\cos \! \left(k_{n} \Delta \right)-1\right) \left(\cos^{2}\left(k_{n} x_{j}\right)\right)-\sin \! \left(k_{n} x_{j}\right) \sin \! \left(k_{n} \Delta \right) \cos \! \left(k_{n} x_{j}\right)-\frac{\cos \left(k_{n} \Delta \right)}{2}+\frac{3}{2}\right) \left(\cos \! \left(k_{n} \Delta \right)-1\right)}
    {\pi j_N \Delta  
    },\nonumber
\end{eqnarray}
and therefore
\begin{eqnarray}
\langle 0\vert \hat{H}_{j}\vert 0 \rangle
 &=&\sum_{n=1}^{j_N}\frac{n}{2j_N \Delta  \,\pi }
 \left(\cos \! \left(\frac{2 \pi n}{j_N}\right)-1\right) \left[\left(-2 \left(\cos \! \left(\frac{2 \pi n}{j_N}\right)-1\right) \left(\cos^{2}\left(\frac{2 \pi  n x_j }{j_N\Delta}\right)\right)\right.\right.\nonumber\\
&& \left.\left.+\cos \! \left(\frac{2 \pi n}{j_N}\right)-3+2 \sin \! \left(\frac{2 \pi  n x_j}{j_N\Delta}\right) \sin \! \left(\frac{2 \pi n}{j_N}\right) \cos \! \left(\frac{2 \pi  n x_j}{j_N\Delta}\right)\right)\right].
\end{eqnarray}

A numerical analysis shows that for large values of $x$ the expectation value goes to a constant depending on $\Delta$, and therefore the integral of the Hamiltonian density in a bounded region of order $l_0 \gg r_S$   can be canceled, after integration in the radial variable, by choosing the deficit angle. For a state with $\Delta$ of order Planck, $\Lambda$ is much smaller than one.
Since the expectation value of the gravitational part is non-vanishing and the true Hamiltonian is the difference between the gravitational and the matter portions,  it is clear that the vacuum for a self gravitating field is not the Minkowskian vacuum we computed above.  The excitation level of the matter portion depends on the gravitational part, which in turn depends on $\sigma$. The minimum eigenvalue of the Hamiltonian may correspond to different vacua, associated with different  excitations of the matter and the gravitational field. For the choice made above, the excitations become smaller as $a$ grows. For example, for $a=2$
\begin{equation}
\langle 0 \vert \hat{H}_{\rm grav}\vert 0\rangle=
\frac{\Delta  \sqrt{2}}{12 \,{l_0} \,\pi^{\frac{3}{2}} \sqrt{1-2 \Lambda}\, \sqrt{G}}
\end{equation}
which should be compared with the first term of (\ref{38}), which was computed for $a=1$. The minimum of the gravitaional energy is never exactly zero, as there is dispersion, and is very small. For instance for $a=1$ it is of the order $E_{\rm Planck}$ and $\sigma\sim \ell_{\rm Planck}^{-2}$. The excitation level of the matter part cannot grow indefinitely since we have a square root of a modulus, a maximum will be reached beyond which one would excite the complete system.  The fundamental state depends on $\sigma$ and for the given family of $a$ given there are less excitations for large $a$'s. The different vacua correspond to different levels of radiation of the system.

\section{Conclusions}

We have studied spherically symmetric loop quantum gravity coupled to a spherically symmetric scalar field deparameterized with a second spherically symmetric scalar field working as a clock. We analyzed the problem asymptotically, far away from the Schwarzschild radius. This allows us to obtain in closed form the eigenvalue equation for the Hamiltonian constraint of the gravitational part, in particular its fundamental state. We then studied perturbatively in powers of Newton's constant the matter part of the Hamiltonian constraint. The total Hamiltonian of the system is a square root of an absolute value of the difference of the gravitational and matter term. This correlates the excitations of gravity with those in the matter. In particular it implies that the fundamental states of gravitating scalar fields are not unique and excitations of the matter fields are allowed and will in general be present in the system. We show that the coupling to gravity provides a natural cutoff and the perturbative theory ends up being formulated on a lattice. 

 The exact treatment of the true Hamiltonian appears quite challenging and may require new techniques we plan to develop in future works that could in particular allow to study the formation of black holes and critical phenomena from a completely quantum mechanical viewpoint

\section{Acknowledgements}
This work was supported in part by Grant NSF-PHY-1903799, NSF-PHY-2206557, funds of the
Hearne Institute for Theoretical Physics, CCT-LSU, Pedeciba, Fondo Clemente Estable
FCE 1 2019 1 155865.


\begin{thebibliography}{9}
\bibitem{usreview}R. Gambini, J. Olmedo, J. Pullin, 
[arXiv:2211.05621 [gr-qc]] to appear in "Handbook of Quantum Gravity", Cosimo Bambi, Leonardo Modesto, Ilya Shapiro (editors), Springer (2023)
\bibitem{choptuik} M.~W.~Choptuik,
Phys. Rev. Lett. \textbf{70}, 9-12 (1993)
doi:10.1103/PhysRevLett.70.9
\bibitem{previous}
R.~Gambini and J.~Pullin,
Class. Quant. Grav. \textbf{40}, no.8, 085016 (2023)
doi:10.1088/1361-6382/acc510
[arXiv:2303.09392 [gr-qc]].
\bibitem{gravityquantized}
M.~Domagala, K.~Giesel, W.~Kaminski and J.~Lewandowski,
Phys. Rev. D \textbf{82}, 104038 (2010)
doi:10.1103/PhysRevD.82.104038
[arXiv:1009.2445 [gr-qc]]; T. Thiemann 
[arXiv:astro-ph/0607380 [astro-ph]]; V.~Husain and T.~Pawlowski,
Phys. Rev. Lett. \textbf{108}, 141301 (2012)
doi:10.1103/PhysRevLett.108.141301
[arXiv:1108.1145 [gr-qc]];
K.~Giesel and T.~Thiemann,
Class. Quant. Grav. \textbf{32}, 135015 (2015)
doi:10.1088/0264-9381/32/13/135015
[arXiv:1206.3807 [gr-qc]].
\bibitem{kuchar}
K.~V.~Kuchar,
Phys. Rev. D \textbf{50}, 3961-3981 (1994)
doi:10.1103/PhysRevD.50.3961
[arXiv:gr-qc/9403003 [gr-qc]].
\bibitem{rastgoo}
R.~Gambini, J.~Pullin and S.~Rastgoo,
Gen. Rel. Grav. \textbf{43}, 3569-3592 (2011)
doi:10.1007/s10714-011-1252-0
[arXiv:1105.0667 [gr-qc]].
\bibitem{GieselThiemann}
K.~Giesel and T.~Thiemann,
Class. Quant. Grav. \textbf{27}, 175009 (2010)
doi:10.1088/0264-9381/27/17/175009
[arXiv:0711.0119 [gr-qc]].

\bibitem{kaminski}
W. Kaminski, J. Lewandowski and M. Bobienski, Class. Quant. Grav. 23, 2761 (2006) doi:10.1088/0264-9381/23/9/001
[gr-qc/0508091]; W. Kaminski, J. Lewandowski and A. Okolow, Class. Quant. Grav. 23, 5547 (2006) doi:10.1088/0264-
9381/23/18/005 [gr-qc/0604112].
\end{thebibliography}
\end{document}